\documentclass[11pt,a4paper]{article}
\usepackage{graphics,graphicx,array,hhline,color,float,mdwlist,amsmath,amsfonts,amssymb,cite,multirow,xcolor,subcaption,hyperref,tabularx,placeins,xspace}
\usepackage[left=2cm,right=2cm,top=2cm,bottom=2cm,head=1cm,foot=1cm]{geometry}

\usepackage{t1enc}
\usepackage[utf8]{inputenc}

\begin{document}
\title{Perturbative accelerating solutions of relativistic hydrodynamics}
\author{B\'alint Kurgyis$^{1}$\footnote{Presented at the 10th Bolyai-Gauss-Lobachevsky Conference by B. K.},
M\'at\'e Csan\'ad$^{1}$\\
$^1$ E\"otv\"os Lor\'and University, H-1117 Budapest, P\'azm\'any P. s. 1/A, Hungary}
\maketitle

%\keyword{relativistic hydrodynamics, solutions, Hubble flow, acceleration}
%\begin{document}

\begin{abstract}
In ultra-relativistic collisions of heavy ions, the strongly interacting Quark Gluon Plasma (sQGP) is created.
The fluid nature of the sQGP was one of the important discoveries of high energy heavy ion physics in the last 
decades. Henceforth the explosion of this matter may be described by hydrodynamical models.
Besides numerical simulations, it is important to study the analytic solutions of the equations of hydrodynamics,
as these enable us to understand the connection of the final and initial states better. In this paper
we present a perturbative, accelerating solution of relativistic hydrodynamics, on top of a known
class of solutions describing Hubble-expansion. We describe the properties of this class of perturbative
solutions, and investigate a few selected solutions in detail.
\end{abstract}

\section{Introduction}
The equations of perfect hydrodynamics have no internal scale, and hence they describe aspects of the time evolution of
systems with vastly different sizes: from galactic clusters and galaxies through stars, planets and human-scale
systems, down to the femtometer scale sQGP, created in heavy ion collisions at
RHIC~\cite{Adcox:2004mh,Adams:2005dq} and the
LHC~\cite{Aamodt:2010jd, Aamodt:2010pa, CMS:2012aa, Chatrchyan:2012ta}.
The sQGP is formed in heavy ion collisions after an initial thermalization time $\mathcal{O}(1\:{\rm fm}/c)$,
its evolutions lasts $\mathcal{O}(10\:{\rm fm}/c)$, after that it creates hadrons in the hadronization.
We observe these hadrons, and hydrodynamics may be used to infer the time evolution and the initial
state from the hadron final state distributions.

Hydrodynamics is based on the local conservation of energy and momentum, expressed through

\begin{align}
\partial_{\nu}T^{\mu\nu} = 0,\label{e:tmunucons}
\end{align}
with $T^{\mu\nu}$ being the energy-momentum tensor. In case of a perfect fluid, this can be written as

\begin{align}
T^{\mu\nu}=(\epsilon + p) u^{\mu}u^{\nu}-pg^{\mu\nu}.
\end{align}
where $u^\mu$ is the flow field (subject to the $u_\mu u^\mu =1$ constraint), $\epsilon$ is the energy density,
$p$ is the pressure and $g^{\mu\nu}$ is the metric tensor, $\rm{diag}(1,-1,-1,-1)$. The Equation of State (EoS) closes this set of equations:

\begin{align}
\epsilon = \kappa p
\end{align}
where $\kappa$ is the EoS parameter, which may depend on the temperature. In this paper we assume
constant values, even if $\kappa(T)$ type of solutions of relativistic hydrodynamics are know~\cite{Csanad:2012hr}.
In case of the above described perfect fluid, continuity for the entropy density $\sigma=(\epsilon+p)/T$ follows from
the above equations, and a similar continuity equation for the density of some conserved charge ($n$) may be prescribed:

\begin{align}
\partial_{\mu} (\sigma u^\mu) = 0,\\
\partial_{\mu} (n u^\mu) = 0.\label{e:cont0}
\end{align}
With this, a solution of the equations is a set of fields $(u^\mu,p,n)$ or $(u^\mu,p,\sigma)$, given in terms of coordinates $x^\mu$,
where sometimes also the coordinate proper-time is introduced as $\tau=\sqrt{x_\mu x^\mu}$, along some
scaling variable $s(x^\mu)$ that describes the spatial profile of the densities in the solution.

The discovery of the fluid nature of the sQGP produced a revival of interest for solutions of hydrodynamics,
beyond the well-known Landau-Khalatnikov~\cite{Landau:1953gs,Khalatnikov:1954aa} and
Hwa-Bjorken~\cite{Hwa:1974gn,Bjorken:1982qr} solutions.
Besides numerical simulations (see e.g. Refs.~\cite{Shen:2014vra,Pang:2016igs,Weller:2017tsr} for recent examples), multiple
advanced analytic solutions were found in the last
decade~\cite{Csorgo:2003ry,Csorgo:2006ax,Nagy:2007xn,Csanad:2012hr,Borshch:2007uf,Pratt:2008jj,Gubser:2010ze,Csanad:2014dpa}.
One important example is the simple, ellipsoidal Hubble-flow described in Ref.~\cite{Csorgo:2003ry}, which
describes hadron and photon observables well~\cite{Csanad:2009wc,Csanad:2011jq}. However, this solution
lacks acceleration, and while Hubble-flow is natural in the final state, initial pressure gradients may be important
in understanding the time evolution of this system. In this paper we attempt to find accelerating
perturbations on top of Hubble-flow.

\section{Perturbative Solutions of Hydrodynamics}

The equation for the conservation of energy and momentum density, Equation~(\ref{e:tmunucons}) may be
projected onto $u^\mu$, producing a Lorentz-parallel and a Lorentz-orthogonal equation,
similarly to Refs.~\cite{Csanad:2012hr,Nagy:2007xn}:

\begin{align}
\kappa u^\mu\partial_\mu p+(\kappa+1)p\partial_\mu u^\mu=0\label{e:energy0}\\
(\kappa+1)pu^\mu\partial_\mu u^\nu=(g^{\mu\nu}-u^\mu u^\nu)\partial_\mu p,\label{e:euler0}
\end{align}
where the first is called the energy equation, and the second is the Euler equation of relativistic hydrodynamics.
If a given solution is given in terms of $(u^\mu,p,n)$, then perturbations on top of this solution may be given as:

\begin{align}
u^\mu &\rightarrow u^\mu + \delta u^\mu,\label{e:pert:u}\\
p &\rightarrow p + \delta p, \label{e:pert:p}\\
n &\rightarrow n + \delta n,\label{e:pert:n}
\end{align}
where we restrict ourselves to a conserved charge here, but the continuity may be understood
for the entropy density just as well. Now if these perturbations are small, then the equations of
hydrodynamics may be given in first order. First of all, the perturbations of the flow field must fulfill

\begin{align}
(u^\mu+\delta u^\mu)(u_\mu+\delta u_\mu)=1
\end{align}
which yields the first order equation of

\begin{align}
u_\mu \delta u^\mu=0.\label{e:orth}
\end{align}
With this, we may substitute the perturbed fields in Equations~(\ref{e:pert:u})--(\ref{e:pert:n})
into the equations of hydrodynamics, Equations~(\ref{e:cont0})--(\ref{e:euler0}).
For the continuity equation, we get the following first order equation:

\begin{align}
u^\mu\partial_\mu\delta n+\delta n\partial_\mu u^\mu+\delta u^\mu\partial_\mu n
+ n\partial_\mu\delta u^\mu = 0.\label{e:cont1}
\end{align}
For the energy equation, we obtain:

\begin{align}
\kappa\delta u^\mu\partial_\mu p+\kappa u^\mu\partial_\mu\delta p+
(\kappa+1)\delta p\partial_\mu u^\mu+(\kappa+1)p\partial_\mu \delta u^\mu=0.\label{e:energy1}
\end{align}
And for the Euler-equation, the first order perturbative equation is

\begin{align}
(\kappa+1)\delta p u^\mu \partial_\mu u^\nu+(\kappa+1) p \delta u^\mu \partial_\mu u^\nu+
(\kappa+1) p u^\mu \partial_\mu \delta u^\nu=(g^{\mu\nu}-u^\mu u^\nu)\partial_\mu \delta p-
\delta u^\mu u^\nu \partial_\mu p-u^\mu \delta u^\nu \partial_\mu p.\label{e:euler1}
\end{align}

To perform a basic consistency check of the above equations, one may investigate what happens when
the basic solution of a fluid at rest. The flow and pressure is then

\begin{align}
u_\mu=(1,0,0,0)\textnormal{ and }p=p_0.
\end{align}
One may immediately observe, that $\partial_\mu u^\mu=0$, $\partial_\mu p=0$, and $u^\mu\partial_\mu=\partial_t$.
With this, the linearized energy and Euler equations become

\begin{align}
\kappa\partial_t\delta p+(\kappa+1)p\partial_\mu\delta u^\mu&=0,\\
(\kappa+1)p\partial_t\delta u^\nu-(u^\mu u^\nu -g^{\mu \nu})\partial_\mu \delta p&=0.
\end{align}
The time derivative of the energy equation is then

\begin{align}
\kappa\partial_t^2\delta p+(\kappa+1)p\partial_t\partial_\mu\delta u^\mu=0.\label{e:energyderivative0}
\end{align}
Let us then introduce the $Q^{\mu\nu}=(u^\mu u^\nu -g^{\mu \nu})$ operator --- which is here
nothing else than ${\rm diag}(0,1,1,1)$. Then the effect of $Q_{\rho\nu}\partial^\rho$ on the Euler equation is

\begin{align}
(\kappa+1)p\partial_t\partial_\nu\delta u^\nu +\Delta\delta p=0,\label{e:EulerQd0}
\end{align}
where we observed that

\begin{align}
Q_{\rho\nu}\partial^\rho Q^{\mu\nu}\partial^\mu= (\partial_x^2 +\partial_y^2 + \partial_z^2)=\Delta.
\end{align}
From Equations~(\ref{e:energyderivative0}) and (\ref{e:EulerQd0}), we obtain

\begin{align}\label{hul}
\partial_t^2\delta p-\frac{1}{\kappa}\Delta\delta p=0,
\end{align}
which means that, as expected, pressure perturbations behave as waves with a speed of sound of
$c_s={1}/{\sqrt{\kappa}}$.

\section{Perturbations on Top of Hubble-Flow}

As mentioned above, in Ref.~\cite{Csorgo:2003ry} a Hubble-type of self-similar solution is given,
with a flow field

\begin{align}
u^\mu = \frac{x^\mu}{\tau},\label{e:Hubbleflow}
\end{align}
where again $\tau^2=x^\mu x_\mu$ is the coordinate proper-time.
The basic quantity of this solution is the scale variable assuring self-similarity, for which
the comoving derivative vanishes:

\begin{align}
u^\mu \partial_\mu S=0.\label{e:scale}
\end{align}
Since in this case, $u^\mu \partial_\mu=\partial_\tau$, the following simple pressure field and density can
be obtained:

\begin{align}
n&=n_0\left(\frac{\tau_0}{\tau}\right)^3\mathcal{N}(S),\label{e:Hubblen}\\ 	
p&=p_0 \left(\frac{\tau_0}{\tau}\right)^{3+\frac{3}{\kappa}},\label{e:Hubblep}\\
\end{align}
where $\mathcal{N}(S)$ is an arbitrary scale function. This solution can be generalized to describe
multipole type of scale variables~\cite{Csanad:2014dpa}, but a standard choice yielding ellipsoidal symmetry is

\begin{align}
S=\frac{x^2}{X^2}+\frac{y^2}{Y^2}+\frac{z^2}{Z^2}
\end{align}
with the coordinates given as $x, y, z$, and the axes of the expanding
ellipsoid are $X, Y, Z$, all linear in time. We will focus here on the spherical case:

\begin{align}
S=\frac{r^2}{\dot R_0^2 t^2},
\end{align}
where $r$ is the radial coordinate, and $\dot R_0$ describes the expansion velocity of the
scale of the solution.  This solution yields the following equations for the perturbations of the fields:

\begin{align}
\delta u^\mu n\frac{\mathcal{N}'}{\mathcal{N}}\partial_\mu S+u^\mu\partial_\mu\delta n +\frac{3\delta n}{\tau}+n\partial_\mu\delta u^\mu&=0\label{e:contH},\\
\kappa u^\mu\partial_\mu \delta p+\frac{3(\kappa+1)}{\tau}\delta p&=-(\kappa+1)p\partial_\mu\delta u^\mu,\label{e:energyH}\\
\frac{\partial_\mu\delta p}{(\kappa+1)p}\left[g^{\mu\nu}-u^\mu u^\nu\right]&=\frac{\kappa-3}{\tau\kappa}\delta u^\nu +u^\mu\partial_\mu\delta u^\nu.\label{e:eulerH}
\end{align}
A similar setup was investigated in Ref.~\cite{Shi:2014kta}, where the authors found expressions for the ripples
propagating on Hubble-flow. Unlike Ref.~\cite{Shi:2014kta}, we will now discuss global perturbations
in terms of $\delta u^\mu$, $\delta p$ and $\delta n$.

In this proceedings paper we do not detail the way this solution was obtained, but simply present the
result for the flow, pressure and density:

\begin{align}
\delta u^\mu&=\delta \cdot F(\tau)
g(x^\mu) \chi (S)\partial^\mu S,\label{e:du1}\\
\delta p&=\delta\cdot p_0\left(\frac{\tau_0}{\tau}\right)^{3+\frac{3}{\kappa}}\pi (S),\label{e:dp1}\\
\delta n&=\delta \cdot n_0\left(\frac{\tau_0}{\tau}\right)^3 h(x^\mu)\nu (S),\label{e:dn1}
\end{align}
where $S$ is the scale variable (with vanishing comoving derivative), $\delta$ is the perturbation scale,
$c$ is an arbitrary constant, $F, h, g$ are profile functions, while $\pi$, $\chi$, $\nu$ are scale functions subject to
the following condition equations:

\begin{align}\label{e:chi:s}
\frac{\chi'(S)}{\chi(S)}&=-\frac{\partial_\mu\partial^\mu S}{\partial_\mu S\partial^\mu S}
-\frac{\partial_\mu S\partial^\mu \ln g(x^\mu)}{\partial_\mu S\partial^\mu S},\\ \label{e:pi:s}
\frac{\pi'(S)}{\chi(S)}&=(\kappa+1)\left[F(\tau)\left(u^\mu\partial_\mu g(x^\mu)-
\frac{3g(x^\mu)}{\kappa\tau}\right)+F'(\tau)g(x^\mu)\right],\\ \label{e:nu:s}
\frac{\nu (S)}{\chi(S)\mathcal{N}'(S)} &=-\frac{F(\tau)g(x^\mu) \partial_\mu S\partial^\mu S}{u^\mu\partial_\mu h(x^\mu)}.
\end{align}
In simple terms, these equations can be translated to the following conditions:
\begin{itemize}
\item The scale variable $S$ fulfills $u_\mu \partial^\mu S=0$ with the original flow field.
\item The right hand sides of Equations~(\ref{e:chi:s})--(\ref{e:nu:s}) depends only on $S$.
\end{itemize}
First of all, let us restrict ourselves to the simplest case of $g(x^\mu)=1$ here, in order to describe the way this
class of perturbative solutions works. This gives a simple form for $F$ as

\begin{align}
F(\tau)=\tau+c\tau_0\left(\frac{\tau}{\tau_0}\right)^\frac{3}{\kappa}.
\end{align}
Then let us select an $h$ function that leads to simpler condition equations:

\begin{align}
h(x^\mu)&=\ln\left(\frac{\tau}{\tau_0}\right)+ \frac{c\kappa}{3-\kappa}\left(\frac{\tau}{\tau_0}\right)^{\frac{3}{\kappa}-1},
(\textnormal{ if } \kappa\neq 3),\label{e:hdef1}\\
h(x^\mu)&=(1+c)\ln\left(\frac{\tau}{\tau_0}\right), (\textnormal{ if }\kappa=3).\label{e:hdef2}
\end{align}
The above choices of  transforms Equations~(\ref{e:chi:s})--(\ref{e:nu:s}) to the simple equations of

\begin{align}
\frac{\chi'(S)}{\chi(S)}&=-\frac{\partial_\mu\partial^\mu S}{\partial_\mu S\partial^\mu S},\label{e:chi:s2}\\
\frac{\pi'(S)}{\chi(S)}&=\frac{(\kappa+1)(\kappa-3)}{\kappa} \label{e:pi:s2}\\
\frac{\nu (S)}{\chi(S)\mathcal{N}'(S)}&=-\tau^2\partial_\mu S\partial^\mu S.\label{e:nu:s2}
\end{align}
While more general solutions can also be found, a broad
class of perturbative solutions can already be given, if suitable $S$ scale variables and associated
$\pi$, $\chi$, $\nu$ and $h$ functions are found. Such suitable scale variables include

\begin{align}
S=\frac{r^m}{t^m}, \qquad S=\frac{r^m}{\tau^m}, \qquad S=\frac{\tau^m}{t^m}.\label{e:scales}
\end{align}
In the next section, we will detail one particular sub-class of this class of solutions.

\section{A Selected Sub-Class of Perturbative Solutions}

If we introduce $h$ as given in Equations~(\ref{e:hdef1})--(\ref{e:hdef2}) and
$S$ as $r^m/t^m$, we obtain the following scale functions:

\begin{align}
\chi(S)&=S^{-\frac{m+1}{m}},\label{e:rntn:chi}\\
\pi(S)&=-\frac{(\kappa+1)(\kappa-3)}{\kappa}m  S^{-\frac{1}{m}},\label{e:rntn:pi}\\
\nu (S)&=m^2  S^{\frac{m-1}{m}}\left(S^\frac{2}{m}-1\right)\left(1-S^{-\frac{2}{m}}\right)\mathcal{N}'(S).\label{e:rntn:nu}
\end{align}
This sub-class of solutions contains an arbitrary parameter $c$, the $\delta$ perturbation scale, the $m$ exponent
and the $\mathcal{N}(S)$ scale function (included in the original Hubble-solution as well).
Let us chose $m=-1$, then the scale functions are

\begin{align}
\chi(S)&=1,\\
\pi(S)&=\frac{(\kappa+1)(\kappa-3)}{\kappa}  S,\\
\nu(S)&=  \left(1-S^2\right)^2\mathcal{N}'(S).
\end{align}
Let us furthermore choose a suitable $\mathcal{N}$, leading to a Gaussian profile:

\begin{align}
\mathcal{N}(S)=e^{-bS^{-2}}=e^{-b\frac{r^2}{t^2}}
\end{align}
With these, the perturbed fields (for $\kappa\neq 3$, the special case of $\kappa=3$ is discussed in Equation~(\ref{e:hdef2})) are as follows:

\begin{align}
\delta u^\mu&=
\delta \cdot \left[\tau+c\tau_0\left(\frac{\tau}{\tau_0}\right)^\frac{3}{\kappa}\right]
\partial^\mu S,\label{e:durm}\\
\delta p&=
\delta\cdot p_0\left(\frac{\tau_0}{\tau}\right)^{3+\frac{3}{\kappa}}\frac{(\kappa+1)(\kappa-3)}{\kappa}S,\label{e:dprm}\\
\delta n&=
\delta \cdot n_0\left(\frac{\tau_0}{\tau}\right)^3 \left[\ln\left(\frac{\tau}{\tau_0}\right)+ c\frac{\kappa}{3-\kappa}\left(\frac{\tau}{\tau_0}\right)^{\frac{3}{\kappa}-1}\right] S^{-3}\left(1-S^2\right)^2 2b\mathcal{N}(S).\label{e:dnrm}
\end{align}
For the visualisation of these fields, let us chose parameter values from Refs.~\cite{Csanad:2009wc,Csanad:2011jq} as
$\tau_0 = 7.7\:{\rm fm}/c$, $\kappa=10$ and $b=-0.1$.

\begin{figure}[H]
\centering
     \includegraphics[width=0.496\linewidth]{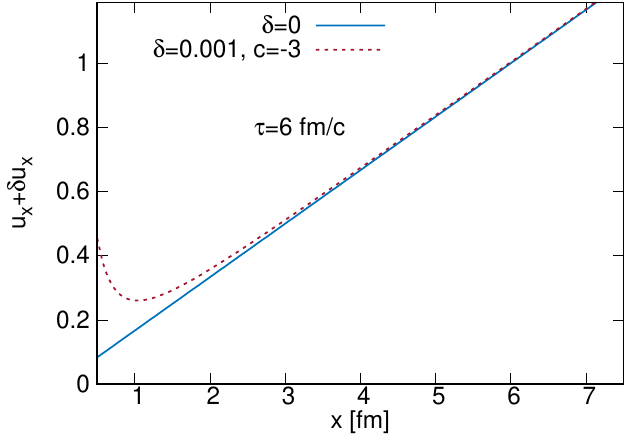}
     \includegraphics[width=0.496\linewidth]{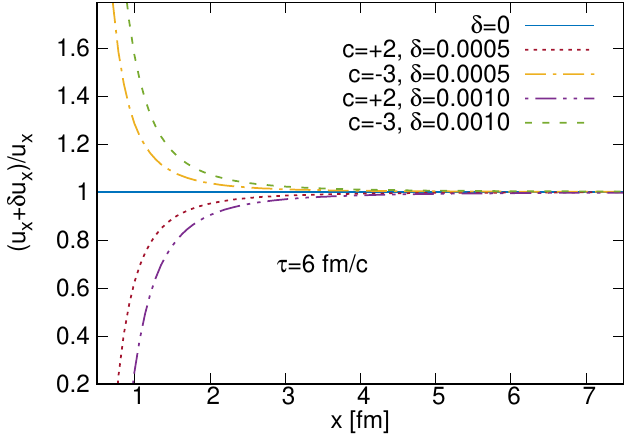}\\
     \hspace{0.02\linewidth}\includegraphics[width=0.975\linewidth]{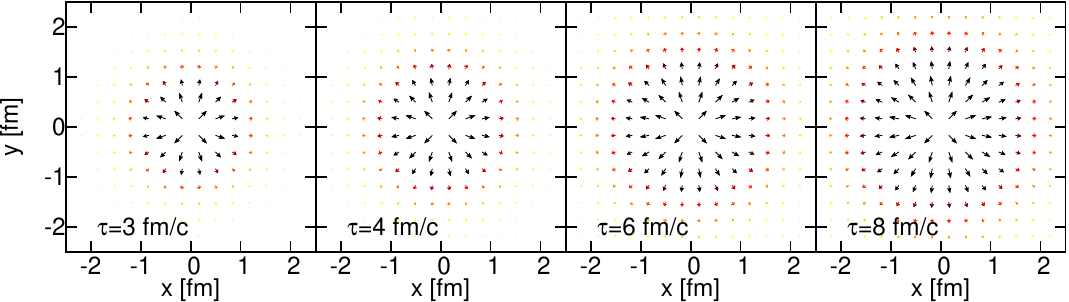}
  \caption{The perturbed flow field component ($u_x+\delta u_x$) is shown in the left plot as a function of $x$,
  for $\tau=6$ fm/$c$ (the other parameters are given in the text). The right plot indicates the relative change
  $(u_x+\delta u_x)/u_x$ for various $\delta$ and $c$ values. The bottom plot shows the flow perturbation field
  $(\delta u_x,\delta u_y)$   in the transverse plane, for various proper-time values.}
  \label{f:du}
\end{figure}

\begin{figure}[H]
\centering
     \includegraphics[width=0.496\linewidth]{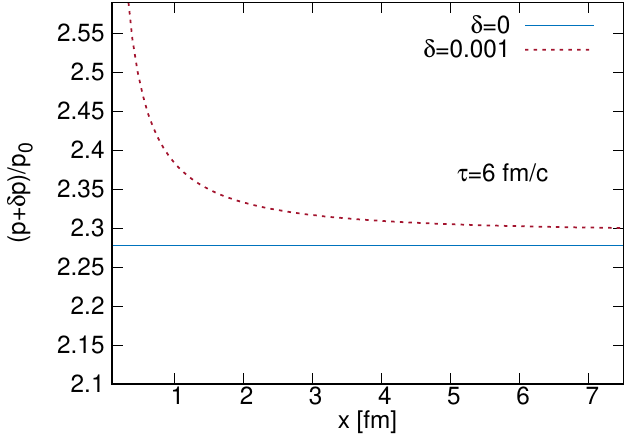}
     \includegraphics[width=0.496\linewidth]{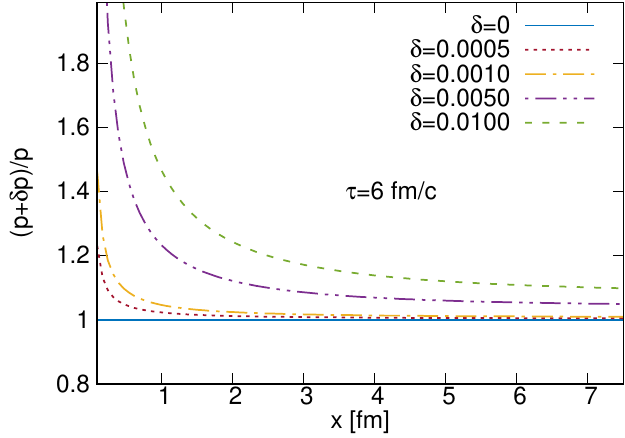}\\
     \hspace{0.03\linewidth}\includegraphics[width=0.965\linewidth]{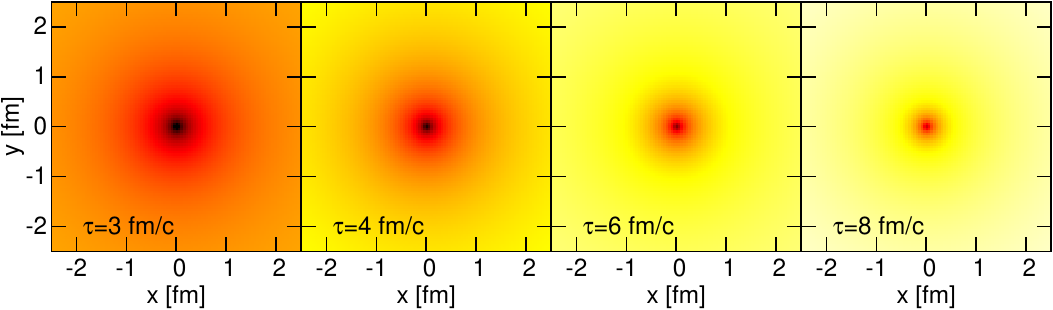}
  \caption{The perturbed pressure $p+\delta p$ is shown in the left plot as a function of $x$,
  for $\tau=6$ fm/$c$ (the other parameters are given in the text). The right plot indicates the relative change
  $(p+\delta p)/p$ for various $\delta$ and $c$ values.  The bottom plot shows the pressure perturbation $\delta p$
  in the transverse plane, for various proper-time values.}
  \label{f:dp}
\end{figure}

\begin{figure}[H]
  \centering
     \includegraphics[width=0.495\linewidth]{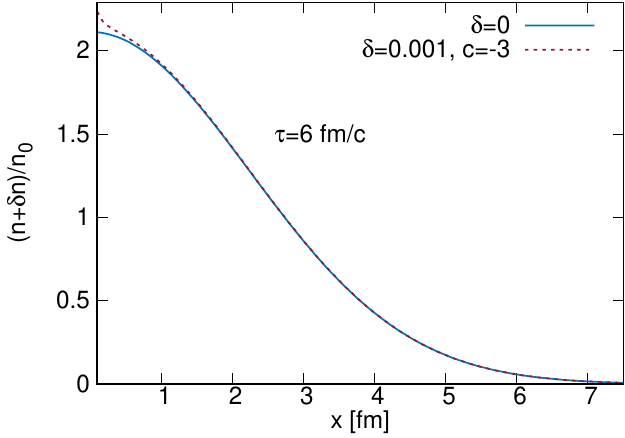}
     \includegraphics[width=0.495\linewidth]{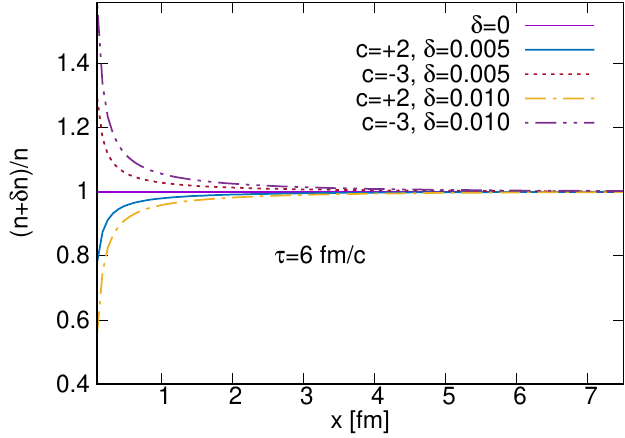}
  \caption{The perturbed density  $n+\delta n$ is shown in the left plot as a function of $x$,
  for $\tau=6$ fm/$c$ (the other parameters are given in the text). The right plot indicates the relative change
  $(n+\delta n)/n$ for various $\delta$ and $c$ values.  }
  \label{f:dn}
\end{figure}

On the top left panel of Figure~\ref{f:du}, a slice of the $x$ component of
the flow field is shown with $\tau=6$ fm/$c$, $c=-3$ and $\delta=0.001$. The perturbation is the most important
in the center, it also changes the direction of the field, but it vanishes for large radial distances.
The top right panel indicates the $c$ and $\delta$ dependence of the relative perturbed fields.
We observe here that for this particular solution, the relative perturbation increases to very large
values for very small distances. The bottom panel of Figure~\ref{f:du} indicates the transverse flow field for various proper-time slices,
showing that also the direction of the flow is perturbed for some particular distances.
Next, let us investigate the pressure perturbation. The top panels of Figure~\ref{f:dp} shows the pressure field with fixed
values of $\delta=0.001$ and $\tau=6$ fm/$c$ (there is no $c$-dependence in $p$).  Again it is clear that the perturbation
vanishes for increasing radial distance, and increases for small distances. It is an important next step to present a
sub-class of perturbative solutions that does not exhibit this feature. One may also note that $\delta$ controls
the perturbation magnitude, as also visible in the ratio plots in the top right panel of Figure~\ref{f:dp}.
On the bottom panel, the time evolution of the
pressure perturbation is given in the transverse plane, showing a vanishing perturbation for large times.
Finally, let us investigate the behavior of the density $n$. The left panel of Figure~\ref{f:dn} (with $\tau=6$ fm/$c$,
$\delta=0.001$ and $c=-3$) indicates again a vanishing perturbation for large distances. The right panel
shows the relative perturbation and its dependence on $\delta$ and $c$. As it can be seen
also in the figures, the perturbations become larger than the original fields for very small radial distances and large $\delta$
values, i.e. the method breaks down in these cases. This sets a limit to the applicability of the particular investigated
perturbative solutions. Note however, that this is not  necessarily a general property of the whole class of solutions.

With these fields at hand, and utilizing a freeze-out hypersurface similarly to e.g., Ref.~\cite{Csanad:2009wc}, one
may evaluate observables such
as transverse momentum distribution, flow and Bose-Einstein correlation radii. We plan to do this
in a subsequent publication.

\section{Conclusions}

In this paper we presented the method of obtaining perturbative solutions of relativistic hydrodynamics on top
of known solutions. A new perturbative class of solutions on top of Hubble flow was discussed, and the
modified fields were investigated in detail. These fields were scaled to a single $\delta$ perturbation parameter,
and several scale functions appeared, subject to condition equations. As a subsequent step, we plan to describe
more particular sub-classes of solutions. We also plan to calculate the modification of observables and in case
of realistic geometries, we plan to compare them to measurements.

\section*{Acknowledgments}
The authors are supported by the New National Excellence program of the Hungarian Ministry of Human Capacities and the NKFIH grant FK-123842. M. Cs. was also supported by the J\'anos Bolyai Research Scholarship of the Hungarian Academy of Sciences.

\bibliographystyle{../../../prlsty}
\bibliography{../../../Master}

\end{document}